\documentclass[sigconf]{acmart}

\usepackage{url}
\usepackage{hyperref}
\usepackage{subcaption}
\usepackage{caption}
\usepackage[breakable,skins]{tcolorbox}
\usepackage{multirow}
\usepackage{enumitem}
\usepackage{float}
\usepackage{lipsum}
\usepackage{wrapfig}
\usepackage{textcomp}
\usepackage{tabularx}
\usepackage{booktabs}
\usepackage{graphicx}

\acmConference[NLBSE 2024]{3rd International Workshop on Natural Language based Software Engineering}{April 2024}{Lisbon, Portugal}

\AtBeginDocument{%
  \providecommand\BibTeX{{%
    \normalfont B\kern-0.5em{\scshape i\kern-0.25em b}\kern-0.8em\TeX}}}

\copyrightyear{2024}
\acmYear{2024}
\setcopyright{acmcopyright}\acmConference[NLBSE 2024]{3rd International Workshop on Natural Language based Software Engineering}{April 2024}{Lisbon, Portugal}
\acmPrice{15.00}
\acmDOI{xxx}
\acmISBN{x}

\begin{document}

\title{Software Vulnerability and Functionality Assessment using LLMs}

\author{Rasmus Ingemann Tuffveson Jensen}
\email{rasmus.jensen@jpmorgan.com}
\affiliation{%
  \institution{JP Morgan AI Research}
  \city{London}
  \country{UK}}

\author{Vali Tawosi}
\email{vali.tawosi@jpmorgan.com}
\affiliation{%
  \institution{JP Morgan AI Research}
  \city{London}
  \country{UK}}

\author{Salwa Alamir}
\email{salwa.alamir@jpmorgan.com}
\affiliation{%
  \institution{JP Morgan AI Research}
  \city{London}
  \country{UK}}

\renewcommand{\shortauthors}{Jensen, Tawosi, Alamir}
\newcommand{\rasmus}[1]{\textcolor{blue}{#1}}

\begin{abstract}
  While code review is central to the software development process, it can be tedious and expensive to carry out.  In this paper, we investigate whether and how Large Language Models (LLMs) can aid with code reviews. Our investigation focuses on two tasks that we argue are fundamental to good reviews: (i) flagging code with security vulnerabilities and (ii) performing software functionality validation, i.e., ensuring that code meets its intended functionality. To test performance on both tasks, we use zero-shot and chain-of-thought prompting to obtain final ``approve or reject'' recommendations. As data, we employ seminal code generation datasets (HumanEval and MBPP) along with expert-written code snippets with security vulnerabilities from the Common Weakness Enumeration (CWE).
  Our experiments consider a mixture of three proprietary models from OpenAI and smaller open-source LLMs. We find that the former outperforms the latter by a large margin. Motivated by promising results, we finally ask our models to provide detailed descriptions of security vulnerabilities. Results show that 36.7\% of LLM-generated descriptions can be associated with true CWE vulnerabilities.
\end{abstract}



\maketitle
\section{Introduction}
Code review is the process whereby software developers analyze if peer contributions are of sufficient quality to be integrated into codebases. The practice reduces bugs, increases code quality, and facilitates knowledge transfer \cite{morales2015code}. Code reviews can, however, be costly and tedious to carry out \cite{czerwonka2015code}. Furthermore, poorly performed reviews may foster a negative work environment, provide a false sense of security, and hinder innovation \cite{Sarker2023}. In an effort to address such problems, previous studies have proposed methods to automate code reviews, so far, though, with modest success \cite{tufano2022using}.

Large language models (LLMs) have recently demonstrated remarkable performance on a variety of tasks, including code generation and question answering \cite{chen2021}. In this paper, we investigate whether and how LLMs can aid with code reviews. As models, we consider a mixture of open-source and proprietary LLMs, including models from the Dolly, Falcon, Llama, and GPT families. Acknowledging that reviews contribute to many things, e.g., knowledge sharing and maintainability, our experiments focus on two tasks: (i) flagging code with security vulnerabilities and (ii) performing software functional validation, i.e., ensuring that code meets its intended functionality.\footnote{We do not aim to generate unit tests (which should be applied before a code review). Rather, we aim to determine if a code snippet meets its intended functionality without executing it.} The tasks motivate our research questions:

\begin{enumerate}
    \item[RQ1.] Can LLMs flag code security vulnerabilities?
    \item[RQ2.] Can LLMs do software functional validation?
    \item[RQ3.] Can LLMs simultaneously flag security vulnerabilities and do software functional validation?
    \item[RQ4.] Can LLMs provide feedback on security vulnerabilities?
\end{enumerate} 

\section{Related Work} \label{sec:Related Literature}
Several studies have investigated how machine learning and natural language processing can support code reviews. The field is related to defect prediction, i.e., predicting if a code snippet contains a bug. Studies vary substantially in their approach and the code granularity being analyzed. They may, e.g., consider stand-alone functions, source code files, or file changes (i.e., ``diffs'' in pull requests).

Li \textit{et al.} \cite{Heng2019} consider classification of triplets made up of change descriptions and snippets of old and new code. The aim is to predict if a triplet is accepted by reviewers. The authors propose a deep learning model utilizing word2vec embeddings and convolution layers, reporting F1 scores from 0.44 to 0.50 on data from five software projects.
Shi \textit{et al.} \cite{shi2019automatic} consider pairs of original and changed source files. Aiming to predict if changes are approved by a reviewer, the authors propose a deep learning model with convolution and LSTM layers. Using data from six projects, the authors report F1 scores ranging from 0.40 to 0.57.
Kim \textit{et al.} \cite{kim2008classifying} aim to classify code changes from 12 projects as ``buggy'' or ``clean.'' The authors use SVM on features that include bag-of-words metrics. Reporting 78\% accuracy, the authors note that their data may contain an inflated number of bugs.  
Lu \textit{et al.} \cite{lu2023llama} propose Llama-Reviewer, a framework to fine-tune LLMs for code reviews in a parameter-efficient manner. Considering code changes, the authors, in particular, report an F1 score around 0.70 trying to predict the necessity of code reviews.

Our paper adds to the existing literature by (i) investigating how LLMs may be used to flag (and describe) security vulnerabilities, not (just) logical errors. Secondly, we explore how LLMs may be used to determine if a piece of code meets its intended functionality without executing it. 

\section{Methodology} \label{sec:Experimental Setup}

In this section, we outline our employed datasets, the LLMs we used for experimentation, and our experimental setup.

\subsection{Datasets} \label{subsec:Datasets}

Our experiments utilize three datasets with Python code snippets at the function level.
For all datasets, we only consider code snippets that contain exactly one function for consistency. Furthermore, we only consider observations\footnote{We use the terms ``code snippet'' and ``observation'' interchangeably.} that contain (i) exactly one docstring (from which we extract each code snippet's intended functionality) or (ii) at most one docstring but a clearly stated programming task in a separate column of the original dataset. Our datasets are:

\begin{enumerate}[leftmargin=\parindent,align=left,labelwidth=\parindent,labelsep=0pt]
    \item[\textbf{1.}] \textbf{HumanEval} \cite{chen2021}, from which we employ 148 observations. The original dataset contains 164 observations. However, we discard 16 observations that do not contain exactly one function and exactly one docstring. 
    \item[\textbf{2.}] \textbf{MBPP} \cite{austin2021program}, from which we employ 476 observations. The original dataset contains 974 observations. However, the authors only designate 500 observations for testing. Furthermore, we discard 24 observations that do not contain exactly one function or contain more than one docstring.
    \item[\textbf{3.}]\textbf{SecurityEval} \cite{SecurityEvalGit}, from which we employ 36 observations with security vulnerabilities per the CWE system \cite{CWE}. The original dataset contains 121 observations. However, we discard (i) 78 web-scraped observations (to avoid data leakage as our LLMs might have seen them during training), (ii) 6 observations that do not contain exactly one function and exactly one docstring, and (iii) a single observation associated with CWE-730, a CWE code actively discouraged from use.
\end{enumerate}

None of our employed observations have, to the best of our knowledge, been seen by our LLMs during training. Furthermore, we only consider author-written observations from SecurityEval (published in late 2022). In relation to security vulnerabilities, we denote observations from SecurityEval as ``dirty'' and observations from HumanEval and MBPP as ``clean.'' While our data only contains around 5\% dirty observations, empirical studies, notably, suggest that security vulnerabilities are rare in real codebases \cite{pei2023code}.

\subsection{Models} \label{subsec:Models}
Our experiments consider nine LLMs; six open-source and three proprietary. The open-source models were chosen from the Hugging Face Open LLM Leaderboard in August 2023, aiming to represent influential models.  Based on resource and hardware constraints, we only consider smaller versions of the models. Unless explicitly stated, all hyper-parameters are kept at default values (similar to Hugging Face deployments). The proprietary models are employed through APIs as of April/May 2023 (before OpenAI released updated model versions in June 2023).

Our open-source models are: Falcon-7b-instruct, Llama-2-7b-chat, Llama-2-13b-chat, Dolly-v2-3b, Dolly-v2-7b, and Dolly-v2-12b \cite{falcon40b,touvron2023llama,DatabricksBlog2023DollyV2}. Our proprietary models are: Text-davinci-003, GPT-3.5-turbo, and GPT-4 \cite{OpenAI_Models}. 

\subsection{Experimental Setup} \label{subsec:Experimental Setup}
For all experiments, we report model accuracy. Our data is highly imbalanced in terms of security vulnerabilities, motivating us to also report F1 scores. Previous studies have shown that LLM outputs can vary substantially given small input changes \cite{jha2023codeattack}. For robustness, we therefore run all experiments (i.e., prompts) 10 times, perturbing code snippets between runs. For each code snippet, we randomly apply one of the following transformations: splitting the longest line, replacing tabs with spaces, replacing frequent variable names with ''xxxx'', converting between CamelCase and snake\_case, and doing nothing. The transformations are heavily inspired by \cite{wang2022recode}, designed to change only the syntax, not functionality, of code snippets. For RQ1 through RQ3, we limit LLM answers to 8 new tokens. For RQ4, we limit answers to 100 new tokens.\footnote{Limiting tokens allows faster inference. However, we must allow a sufficient number of tokens to ensure that our models can produce responses as desired.} Below, we address each of our research questions in turn, describing our experimental setup for each. 

\renewcommand{\paragraph}[1]{\vspace{0.5em}\noindent\textbf{#1}}
\paragraph{RQ1.}
We employ zero-shot, binary classification to answer RQ1. As positive observations, we use dirty code snippets (i.e., observations from SecurityEval). As negative observations, we use clean code snippets (i.e., observations from MBPP and HumanEval). To make predictions, we use the prompt in Figure \ref{fig:security_prompt}. An answer (cast to be lowercase) that contains the word ``yes'' is coded as a positive prediction; otherwise, it is coded as a negative prediction.

\setlength{\abovecaptionskip}{1pt}

\begin{figure}[t!]
    \centering
    \begin{tcolorbox}[standard jigsaw, width=0.48\textwidth, colframe=black,opacityback=0.80,boxrule=0.5pt,arc=0pt,outer arc=0pt] \footnotesize
    \raggedright Review the provided code for potential security vulnerabilities. After your review, assign the code a label from [Yes, No]. `Yes' implies that there are security vulnerabilities. `No' implies that there are no security vulnerabilities. Return only the label. \\
    \textbf{Code:} $[$code\_snippet$]$\\
    \textbf{Label: }
    \end{tcolorbox} 
    \caption{RQ1 Zero-shot prompt to flag security vulnerabilities.}
    \label{fig:security_prompt}
\end{figure}

\paragraph{RQ2.}
We employ zero-shot, binary classification to answer RQ2. To this end, we first extract the programming tasks associated with all code snippets. For MBPP, tasks are given in a column in the dataset. For HumanEval and SecurityEval, tasks are given in each code snippet's docstring. To construct positive observations, we pair each code snippet with its correct task description. To construct negative observations, we pair each code snippet with a ``wrong but similar'' description. To this end. we embed all descriptions with OpenAI's Ada-002 model and pair every code snippet with the nearest neighbour to its correct task description (treating each dataset independently).
The idea is to make our classification problem as difficult as possible. Our setup doubles the number of observations in every dataset, making our problem balanced by construction. To classify observations, we use the zero-shot prompt in Figure \ref{fig:verification_prompt}. Predictions are coded as in RQ1.

\begin{figure}[t!]
    \centering
    \begin{tcolorbox}[standard jigsaw, width=0.48\textwidth, colframe=black, before upper={\parindent15pt\noindent},opacityback=0.80,boxrule=0.5pt,arc=0pt,outer arc=0pt] \footnotesize
    \raggedright  Review the provided code and verify that it meets its intended functionality. After your review, assign the code a label from [Yes, No]. `Yes' implies that the code meets its intended functionality. `No' implies that the code does not meet its intended functionality. Return only the label. \\
    \textbf{Code:} $[$code\_snippet$]$\\
    \textbf{Intended functionality:} $[$task\_description$]$\\
    \textbf{Label:}
    \end{tcolorbox}
    \caption{RQ2 Zero-shot prompt for software functional validation.}
    \label{fig:verification_prompt}
\end{figure}

\paragraph{RQ3.} We consider both zero-shot and chain-of-thought prompting to answer RQ3, aiming to produce final ``approve/reject'' recommendations. We employ the same experimental data as in RQ2 (i.e., datasets with correct and wrong task descriptions). For the zero-shot approach, we use the prompt in Figure \ref{fig:chain-of-thought_recommendation_prompt} without the bold italicised text. 
For the chain-of-thought approach, we use the prompt in Figure \ref{fig:chain-of-thought_recommendation_prompt} with the bold italicised text, utilizing answers obtained by the prompts in Figures \ref{fig:security_prompt} and \ref{fig:verification_prompt}. Our setup runs all prompts in sequence, not mixing results between runs. For HumanEval and MBPP, half of our observations are positive, being clean and associated with correct task descriptions, while the other half are negative, being associated with wrong task descriptions even though they are clean. For SecurityEval, all observations are negative, being dirty regardless of their task descriptions. An LLM answer that contains the word ``approve'' is coded as a positive prediction (and otherwise as negative).  

\paragraph{RQ4.}
To investigate if our LLMs can provide specific feedback on security vulnerabilities, we employ what may be described as multi-class, zero-shot classification. First, we ask our LLMs to describe security vulnerabilities in our 36 code snippets from SecurityEval. Our prompt is given in Figure \ref{fig:describe_vulnerabilities}. To evaluate the quality of the generated descriptions, we (i) embed them using openAI's Ada-002 model, (ii) embed 958 true vulnerability names from the CWE system, and (iii) measure how similar each LLM-generated description is to all true vulnerability names using cosine similarity.

\noindent We consider an LLM-generated description to be good if its closest CWE name corresponds to (i) the actual vulnerability in an underlying code snippet or (ii) an immediate relation (e.g., parent, child, or peer) to the former. We note that the 958 CWE vulnerabilities from which we consider true vulnerability names have 3.09 relations on average. The vulnerabilities in our employed data from SecurityEval have 6.39 relations on average.

\begin{figure}[t!]
    \centering
    \begin{tcolorbox}[standard jigsaw, width=0.48\textwidth, colframe=black, before upper={\parindent15pt\noindent},opacityback=0.80,boxrule=0.5pt,arc=0pt,outer arc=0pt] \footnotesize
    \raggedright Review the provided code for potential security vulnerabilities and verify that it meets its intended functionality.  \\
    \textit{\textbf{When asked if the code contains a security vulnerability, a previous review yielded: [answer\_from\_RQ1]. \\
    When asked if the code meets its intended functionality, a previous review yielded: [answer\_from\_RQ2].}} \\
    After your review, assign the code a label from [Approve, Reject]. `Approve' implies that there are no security vulnerabilities and that the code meets its intended functionality. `Reject' implies that there are security vulnerabilities or that the code does not meet its intended functionality. Return only the label.  \\
    \textbf{Code:} [code\_snippet]\\
    \textbf{Intended functionality: }[task\_description]\\
    \textbf{Label:}
    \end{tcolorbox} 
    \caption{RQ3 Zero-Shot and Chain-of-thought prompt to obtain final ``approve or reject'' recommendations.}
    \label{fig:chain-of-thought_recommendation_prompt}
\end{figure}

\begin{figure}[t!]
    \centering
    \begin{tcolorbox}[standard jigsaw, width=0.48\textwidth, colframe=black, before upper={\parindent1pt\noindent},opacityback=0.80,boxrule=0.5pt,arc=0pt,outer arc=0pt] \footnotesize
    
    \raggedright Review the provided code for potential security vulnerabilities. Provide a brief description of any vulnerability you find. Return only the description. \\
    \textbf{Code:} [code\_snippet]\\
    \textbf{Description:}
    \end{tcolorbox} 
    \caption{RQ4 Zero-shot prompt used to obtain descriptions of security vulnerabilities.}
    \label{fig:describe_vulnerabilities}
\end{figure}

\section{Results} \label{sec:Results}

\begin{table*}[t!]
  \caption{Performance on RQ1-RQ4. Metrics averaged over 10 runs. Standard deviations in parentheses. }
  \label{tab:zero-shot_security_vulnerabilities} 
  \centering
  \resizebox{\textwidth}{!}{
  \begin{tabular}{l | r r | r r | r r | r r | r} \hline
     \multirow{2}{*}{LLM}& \multicolumn{2}{c}{\textbf{RQ1}} & \multicolumn{2}{c}{\textbf{RQ2}}& \multicolumn{2}{c}{\textbf{RQ3 (Zero-shot)}}& \multicolumn{2}{c}{\textbf{RQ3 (Chain-of-thought)}} & \multicolumn{1}{c}{\textbf{RQ4}}\\
     & \textbf{Accuracy} & \textbf{F1 {\footnotesize(Dirty)}} & \textbf{Accuracy} & \textbf{F1 {\footnotesize(True Task)}} & \textbf{Accuracy} & \textbf{F1 {\footnotesize(Approve)}}& \textbf{Accuracy} & \textbf{F1 {\footnotesize(Approve)}} & \textbf{Match Rate}\\
    \hline \hline
GPT-4 &  74.7 (1.0) & 29.3 (0.9) & \textbf{88.7 (0.3)} & \textbf{88.2 (0.4)} & \textbf{80.8 (0.5)} & \textbf{76.6 (0.7)} & \textbf{87.2 (0.2)} & \textbf{85.7 (0.3)} & \textbf{36.7 (3.4)} \\
GPT-3.5-turbo & 82.2 (1.0) & 16.2 (3.8) & 57.9 (0.5) & 70.1 (0.3) & 50.5 (0.5) & 65.5 (0.2) & 55.8 (0.7) & 67.4 (0.4) & 20.3 (4.9) \\
Text-davinci-003 & \textbf{95.6 (0.1)} & \textbf{37.9 (3.0)} & 82.9 (0.4) & 84.4 (0.3) & 68.5 (0.3) & 74.7 (0.2) & 80.6 (0.4) & 81.6 (0.3) & 24.4 (6.1) \\
Llama-2-13b-chat-hf & 50.6 (2.3) & \phantom{0}9.3 (1.2) & 51.6 (1.2) & 53.8 (1.5) & 48.1 (0.9) & 53.8 (1.1) & 50.3 (1.6) & 50.2 (1.5) & 19.2 (4.8) \\
Dolly-v2-12b & 54.9 (2.9) & \phantom{0}8.9 (1.8) & 49.2 (0.6) & 59.0 (0.5) & 51.3 (1.5) & 39.8 (1.6) & 48.3 (1.2) & 60.4 (0.8) & 19.2 (3.1) \\
Falcon-7b-instruct & 42.0 (1.1) & \phantom{0}9.8 (0.9) & 50.3 (0.5) & 65.3 (0.5) & 47.7 (0.6) & 61.5 (0.5) & 49.5 (1.6) & 55.2 (1.2) & 19.4 (7.0) \\
Dolly-v2-7b & \phantom{0}9.7 (0.7) & 10.0 (0.3) & 50.0 (0.9) & 64.6 (0.7) & 47.2 (0.6) & 62.0 (0.5) & 47.7 (0.7) & 62.5 (0.4) & 16.7 (6.0) \\
Llama-2-7b-chat-hf & 63.0 (1.3) & 10.9 (2.1) & 49.6 (1.4) & 51.4 (1.2) & 50.7 (1.4) & 57.5 (1.4) & 51.3 (1.7) & 46.8 (2.1) & 17.2 (6.6) \\
Dolly-v2-3b & 22.5 (1.6) & 10.0 (1.0) & 50.2 (0.7) & 63.3 (0.6) & 48.1 (1.1) & 53.7 (1.1) & 48.0 (1.0) & 57.2 (1.0) & 14.7 (4.3) \\
    \hline
  \end{tabular}
  }
\end{table*}

\noindent Our results are displayed in Table \ref{tab:zero-shot_security_vulnerabilities}. Note that the table displays results over our combined dataset.

For RQ1, Text-davinci-003 performs best with an accuracy of 95.6\% and F1 score of 37.9\%.\footnote{ One might be surprised that Text-davinci-003 performs better than GPT-3.5-turbo and GPT-4. We speculate that this might be due to the models' alignment. } All open-source models perform poorly with F1 scores close to (and even below) what one would expect from a predictor assigning the same label to any observation. 

For RQ2, our datasets are balanced by construction. Here, all open-source models consistently achieve accuracy scores around 50\%, similar to predictors assigning the same label to all observations. Among the GPT-models, GPT-4 performs best with an accuracy of 88.7\% and F1 score of 88.2\%.

For RQ3, the open-source models perform poorly regardless of prompting technique. In the proprietary models, we observe substantial performance increases when switching to chain-of-thought prompting. Thus, chain-of-thought prompting appears to increase performances for the proprietary models. Our best-performing model, GPT-4, specifically, sees an increase in its accuracy from 80.8\% to 87.2\% and an increase in F1 from 76.6\% to 85.7\%. Running the same experiments without code perturbations, we obtained similar results, indicating robustness to the perturbations.

For RQ4, the proprietary models appear to outperform the open-source models. Our best-performing proprietary model, GPT-4, in particular, generates vulnerability descriptions that on average can be associated with true CWE names 36.7\% of the time. Considering each model family (GPT, Llama, and Dolly), we note that model performances generally appear to increase with model size.

\section{Threats to Validity} \label{sec:Threats to Validity}

Regarding internal validity, we note that our setup for RQ2 implicitly assumes that all programming tasks in each dataset are unique, i.e., attempt to obtain different things. While we believe this to be the case for HumanEval, it is not given for MBPP and SecurityEval. We also note how LLM answers (i) are stochastic in nature and (ii) can vary given small input changes. To mitigate both effects, we perform perturbations and multiple experimental runs.

Regarding external validity, we stress that our experiments only consider smaller versions of open-source models and code snippets with single functions. Thus, our conclusions do not extend to larger open-source models or source code files; being possible directions for future work.  In future work, our experiments may also be extended to consider specialized open-source models (e.g., Code Llama \cite{roziere2023code}), larger datasets (e.g., EvalPlus \cite{evalplus}), few-shot prompting (see, e.g., \cite{tawosi2023search}), and make comparisons with methods using unit tests or static code analyzers.

\section{Conclusion and Discussion} \label{sec:Discussion and Conclusion}

We have developed an experimental framework to investigate how LLMs can aid in code reviews. Our results show that smaller open-source models generally perform on par with random or dummy classifiers. However, when used to flag security vulnerabilities, the best proprietary model achieves an accuracy of over 95.6\% and an F1 score over 37.9\%. When used to perform software functionality validation, the model achieves an accuracy and F1 score over 88.2\%. Furthermore, vulnerability descriptions from the model can be matched to true vulnerabilities over 36.7\% of the time.

\section*{Disclaimer} 
This paper was prepared for informational purposes by the Artificial Intelligence Research group of JPMorgan Chase \& Co and its affiliates (“JP Morgan”), and is not a product of the Research Department of JP Morgan. JP Morgan makes no representation and warranty whatsoever and disclaims all liability, for the completeness, accuracy or reliability of the information contained herein. This document is not intended as investment research or investment advice, or a recommendation, offer or solicitation for the purchase or sale of any security, financial instrument, financial product or service, or to be used in any way for evaluating the merits of participating in any transaction, and shall not constitute a solicitation under any jurisdiction or to any person, if such solicitation under such jurisdiction or to such person would be unlawful.

\begin{acks}
We are grateful to Ran Zmigrod for discussions and feedback.  
\end{acks}

\bibliographystyle{ACM-Reference-Format}
\bibliography{refs}


\begin{thebibliography}{22}


\ifx \showCODEN    \undefined \def \showCODEN     #1{\unskip}     \fi
\ifx \showDOI      \undefined \def \showDOI       #1{#1}\fi
\ifx \showISBNx    \undefined \def \showISBNx     #1{\unskip}     \fi
\ifx \showISBNxiii \undefined \def \showISBNxiii  #1{\unskip}     \fi
\ifx \showISSN     \undefined \def \showISSN      #1{\unskip}     \fi
\ifx \showLCCN     \undefined \def \showLCCN      #1{\unskip}     \fi
\ifx \shownote     \undefined \def \shownote      #1{#1}          \fi
\ifx \showarticletitle \undefined \def \showarticletitle #1{#1}   \fi
\ifx \showURL      \undefined \def \showURL       {\relax}        \fi
\providecommand\bibfield[2]{#2}
\providecommand\bibinfo[2]{#2}
\providecommand\natexlab[1]{#1}
\providecommand\showeprint[2][]{arXiv:#2}

\bibitem[Almazrouei et~al\mbox{.}(2023)]%
        {falcon40b}
\bibfield{author}{\bibinfo{person}{Ebtesam Almazrouei}, \bibinfo{person}{Hamza
  Alobeidli}, \bibinfo{person}{Abdulaziz Alshamsi}, \bibinfo{person}{Alessandro
  Cappelli}, \bibinfo{person}{Ruxandra Cojocaru}, \bibinfo{person}{Merouane
  Debbah}, \bibinfo{person}{Etienne Goffinet}, \bibinfo{person}{Daniel Heslow},
  \bibinfo{person}{Julien Launay}, \bibinfo{person}{Quentin Malartic},
  \bibinfo{person}{Badreddine Noune}, \bibinfo{person}{Baptiste Pannier}, {and}
  \bibinfo{person}{Guilherme Penedo}.} \bibinfo{year}{2023}\natexlab{}.
\newblock \bibinfo{title}{{Falcon-40B}: {A}n open large language model with
  state-of-the-art performance}.
\newblock
\newblock
\urldef\tempurl%
\url{https://huggingface.co/tiiuae/falcon-7b}
\showURL{%
\tempurl}


\bibitem[Austin et~al\mbox{.}(2021)]%
        {austin2021program}
\bibfield{author}{\bibinfo{person}{Jacob Austin}, \bibinfo{person}{Augustus
  Odena}, \bibinfo{person}{Maxwell Nye}, \bibinfo{person}{Maarten Bosma},
  \bibinfo{person}{Henryk Michalewski}, \bibinfo{person}{David Dohan},
  \bibinfo{person}{Ellen Jiang}, \bibinfo{person}{Carrie Cai},
  \bibinfo{person}{Michael Terry}, \bibinfo{person}{Quoc Le}, {et~al\mbox{.}}}
  \bibinfo{year}{2021}\natexlab{}.
\newblock \bibinfo{title}{Program synthesis with large language models}.
\newblock
\newblock
\showeprint{2108.07732}


\bibitem[Chen et~al\mbox{.}(2021)]%
        {chen2021}
\bibfield{author}{\bibinfo{person}{Mark Chen}, \bibinfo{person}{Jerry Tworek},
  \bibinfo{person}{Heewoo Jun}, \bibinfo{person}{Qiming Yuan},
  \bibinfo{person}{Henrique Ponde de~Oliveira Pinto}, \bibinfo{person}{Jared
  Kaplan}, \bibinfo{person}{Harri Edwards}, \bibinfo{person}{Yuri Burda},
  \bibinfo{person}{Nicholas Joseph}, \bibinfo{person}{Greg Brockman},
  {et~al\mbox{.}}} \bibinfo{year}{2021}\natexlab{}.
\newblock \bibinfo{title}{Evaluating large language models trained on code}.
\newblock
\newblock
\showeprint{2107.03374}


\bibitem[Conover et~al\mbox{.}(2023)]%
        {DatabricksBlog2023DollyV2}
\bibfield{author}{\bibinfo{person}{Mike Conover}, \bibinfo{person}{Matt Hayes},
  \bibinfo{person}{Ankit Mathur}, \bibinfo{person}{Jianwei Xie},
  \bibinfo{person}{Jun Wan}, \bibinfo{person}{Sam Shah}, \bibinfo{person}{Ali
  Ghodsi}, \bibinfo{person}{Patrick Wendell}, \bibinfo{person}{Matei Zaharia},
  {and} \bibinfo{person}{Reynold Xin}.} \bibinfo{year}{2023}\natexlab{}.
\newblock \bibinfo{title}{Free Dolly: Introducing the World's First Truly Open
  Instruction-Tuned LLM}.
\newblock
\newblock
\urldef\tempurl%
\url{www.databricks.com/blog/2023/04/12/dolly-first-open-commercially-viable-instruction-tuned-llm}
\showURL{%
\tempurl}


\bibitem[Czerwonka et~al\mbox{.}(2015)]%
        {czerwonka2015code}
\bibfield{author}{\bibinfo{person}{Jacek Czerwonka}, \bibinfo{person}{Michaela
  Greiler}, {and} \bibinfo{person}{Jack Tilford}.}
  \bibinfo{year}{2015}\natexlab{}.
\newblock \showarticletitle{Code reviews do not find bugs. how the current code
  review best practice slows us down}. In \bibinfo{booktitle}{\emph{37th
  International Conference on Software Engineering}}, Vol.~\bibinfo{volume}{2}.
  \bibinfo{publisher}{IEEE}, \bibinfo{address}{Florence, Italy},
  \bibinfo{pages}{27--28}.
\newblock


\bibitem[Jha and Reddy(2023)]%
        {jha2023codeattack}
\bibfield{author}{\bibinfo{person}{Akshita Jha} {and}
  \bibinfo{person}{Chandan~K Reddy}.} \bibinfo{year}{2023}\natexlab{}.
\newblock \bibinfo{title}{Codeattack: Code-based adversarial attacks for
  pre-trained programming language models}.
\newblock
\newblock
\showeprint{2206.00052}


\bibitem[Kim et~al\mbox{.}(2008)]%
        {kim2008classifying}
\bibfield{author}{\bibinfo{person}{Sunghun Kim}, \bibinfo{person}{James
  Whitehead}, {and} \bibinfo{person}{Yi Zhang}.}
  \bibinfo{year}{2008}\natexlab{}.
\newblock \showarticletitle{Classifying software changes: Clean or buggy?}
\newblock \bibinfo{journal}{\emph{IEEE Transactions on software engineering}}
  \bibinfo{volume}{34}, \bibinfo{number}{2} (\bibinfo{year}{2008}),
  \bibinfo{pages}{181--196}.
\newblock


\bibitem[Li et~al\mbox{.}(2019)]%
        {Heng2019}
\bibfield{author}{\bibinfo{person}{Heng-Yi Li}, \bibinfo{person}{Shu-Ting Shi},
  \bibinfo{person}{Ferdian Thung}, \bibinfo{person}{Xuan Huo},
  \bibinfo{person}{Bowen Xu}, \bibinfo{person}{Ming Li}, {and}
  \bibinfo{person}{David Lo}.} \bibinfo{year}{2019}\natexlab{}.
\newblock \showarticletitle{DeepReview: Automatic Code Review Using Deep
  Multi-instance Learning}. In \bibinfo{booktitle}{\emph{Advances in Knowledge
  Discovery and Data Mining}}, \bibfield{editor}{\bibinfo{person}{Qiang Yang},
  \bibinfo{person}{Zhi-Hua Zhou}, \bibinfo{person}{Zhiguo Gong},
  \bibinfo{person}{Min-Ling Zhang}, {and} \bibinfo{person}{Sheng-Jun Huang}}
  (Eds.). \bibinfo{publisher}{Springer International Publishing},
  \bibinfo{address}{Cham}, \bibinfo{pages}{318--330}.
\newblock
\showISBNx{978-3-030-16145-3}


\bibitem[Liu et~al\mbox{.}(2023)]%
        {evalplus}
\bibfield{author}{\bibinfo{person}{Jiawei Liu}, \bibinfo{person}{Chunqiu Xia},
  \bibinfo{person}{Yuyao Wang}, {and} \bibinfo{person}{Lingming Zhang}.}
  \bibinfo{year}{2023}\natexlab{}.
\newblock \bibinfo{title}{Is Your Code Generated by Chat{GPT} Really Correct?
  Rigorous Evaluation of Large Language Models for Code Generation}.
\newblock
\newblock
\urldef\tempurl%
\url{https://openreview.net/forum?id=1qvx610Cu7}
\showURL{%
\tempurl}


\bibitem[Lu et~al\mbox{.}(2023)]%
        {lu2023llama}
\bibfield{author}{\bibinfo{person}{Junyi Lu}, \bibinfo{person}{Lei Yu},
  \bibinfo{person}{Xiaojia Li}, \bibinfo{person}{Li Yang}, {and}
  \bibinfo{person}{Chun Zuo}.} \bibinfo{year}{2023}\natexlab{}.
\newblock \showarticletitle{LLaMA-Reviewer: Advancing Code Review Automation
  with Large Language Models through Parameter-Efficient Fine-Tuning}. In
  \bibinfo{booktitle}{\emph{IEEE 34th International Symposium on Software
  Reliability Engineering}}. \bibinfo{publisher}{IEEE},
  \bibinfo{address}{Florence, Italy}, \bibinfo{pages}{647--658}.
\newblock


\bibitem[MITRE(2023)]%
        {CWE}
\bibfield{author}{\bibinfo{person}{MITRE}.} \bibinfo{year}{2023}\natexlab{}.
\newblock \bibinfo{title}{Common Weakness Enumeration}.
\newblock
\newblock
\urldef\tempurl%
\url{https://cwe.mitre.org/}
\showURL{%
\tempurl}


\bibitem[Morales et~al\mbox{.}(2015)]%
        {morales2015code}
\bibfield{author}{\bibinfo{person}{Rodrigo Morales}, \bibinfo{person}{Shane
  McIntosh}, {and} \bibinfo{person}{Foutse Khomh}.}
  \bibinfo{year}{2015}\natexlab{}.
\newblock \showarticletitle{Do code review practices impact design quality? a
  case study of the qt, vtk, and itk projects}. In
  \bibinfo{booktitle}{\emph{22nd international conference on software analysis,
  evolution, and reengineering (SANER)}}. \bibinfo{publisher}{IEEE},
  \bibinfo{address}{Montreal, Canada}, \bibinfo{pages}{171--180}.
\newblock


\bibitem[{OpenAI}(2023)]%
        {OpenAI_Models}
\bibfield{author}{\bibinfo{person}{{OpenAI}}.} \bibinfo{year}{2023}\natexlab{}.
\newblock \bibinfo{title}{Models}.
\newblock
\newblock
\urldef\tempurl%
\url{https://platform.openai.com/docs/models}
\showURL{%
\tempurl}


\bibitem[Pei et~al\mbox{.}(2023)]%
        {pei2023code}
\bibfield{author}{\bibinfo{person}{Yulong Pei}, \bibinfo{person}{Salwa Alamir},
  \bibinfo{person}{Rares Dolga}, {and} \bibinfo{person}{Sameena Shah}.}
  \bibinfo{year}{2023}\natexlab{}.
\newblock \showarticletitle{Code Revert Prediction with Graph Neural Networks:
  A Case Study at J.P. Morgan Chase}. In \bibinfo{booktitle}{\emph{1st
  International Workshop on Software Defect Datasets}}.
  \bibinfo{publisher}{ACM}, \bibinfo{address}{San Francisco, CA, USA},
  \bibinfo{pages}{1–5}.
\newblock


\bibitem[Roziere et~al\mbox{.}(2023)]%
        {roziere2023code}
\bibfield{author}{\bibinfo{person}{Baptiste Roziere}, \bibinfo{person}{Jonas
  Gehring}, \bibinfo{person}{Fabian Gloeckle}, \bibinfo{person}{Sten Sootla},
  \bibinfo{person}{Itai Gat}, \bibinfo{person}{Xiaoqing~Ellen Tan},
  \bibinfo{person}{Yossi Adi}, \bibinfo{person}{Jingyu Liu},
  \bibinfo{person}{Tal Remez}, \bibinfo{person}{J{\'e}r{\'e}my Rapin},
  {et~al\mbox{.}}} \bibinfo{year}{2023}\natexlab{}.
\newblock \bibinfo{title}{Code llama: Open foundation models for code}.
\newblock
\newblock
\showeprint{2308.12950}


\bibitem[Sarker et~al\mbox{.}(2023)]%
        {Sarker2023}
\bibfield{author}{\bibinfo{person}{Jaydeb Sarker}, \bibinfo{person}{Asif~Kamal
  Turzo}, \bibinfo{person}{Ming Dong}, {and} \bibinfo{person}{Amiangshu Bosu}.}
  \bibinfo{year}{2023}\natexlab{}.
\newblock \bibinfo{title}{Automated Identification of Toxic Code Reviews Using
  ToxiCR}.
\newblock
\newblock
\showeprint{2202.13056}


\bibitem[Shi et~al\mbox{.}(2019)]%
        {shi2019automatic}
\bibfield{author}{\bibinfo{person}{Shu-Ting Shi}, \bibinfo{person}{Ming Li},
  \bibinfo{person}{David Lo}, \bibinfo{person}{Ferdian Thung}, {and}
  \bibinfo{person}{Xuan Huo}.} \bibinfo{year}{2019}\natexlab{}.
\newblock \showarticletitle{Automatic code review by learning the revision of
  source code}. In \bibinfo{booktitle}{\emph{Proceedings of the AAAI Conference
  on Artificial Intelligence}}, Vol.~\bibinfo{volume}{33}.
  \bibinfo{publisher}{AAAI}, \bibinfo{address}{Honolulu, HI, USA},
  \bibinfo{pages}{4910--4917}.
\newblock


\bibitem[Sunny and Santos(2023)]%
        {SecurityEvalGit}
\bibfield{author}{\bibinfo{person}{Latif Sunny} {and} \bibinfo{person}{Joanna
  Santos}.} \bibinfo{year}{2023}\natexlab{}.
\newblock \bibinfo{title}{SecurityEval}.
\newblock
\newblock
\urldef\tempurl%
\url{www.github.com/s2e-lab/SecurityEval}
\showURL{%
\tempurl}


\bibitem[Tawosi et~al\mbox{.}(2023)]%
        {tawosi2023search}
\bibfield{author}{\bibinfo{person}{Vali Tawosi}, \bibinfo{person}{Salwa
  Alamir}, {and} \bibinfo{person}{Xiaomo Liu}.}
  \bibinfo{year}{2023}\natexlab{}.
\newblock \showarticletitle{Search-Based Optimisation of LLM Learning Shots for
  Story Point Estimation}. In \bibinfo{booktitle}{\emph{International Symposium
  on Search-Based Software Engineering}}. \bibinfo{publisher}{Springer},
  \bibinfo{address}{San Francisco, CA, USA}, \bibinfo{pages}{123--129}.
\newblock


\bibitem[Touvron et~al\mbox{.}(2023)]%
        {touvron2023llama}
\bibfield{author}{\bibinfo{person}{Hugo Touvron}, \bibinfo{person}{Louis
  Martin}, \bibinfo{person}{Kevin Stone}, \bibinfo{person}{Peter Albert},
  \bibinfo{person}{Amjad Almahairi}, \bibinfo{person}{Yasmine Babaei},
  \bibinfo{person}{Nikolay Bashlykov}, \bibinfo{person}{Soumya Batra},
  \bibinfo{person}{Prajjwal Bhargava}, \bibinfo{person}{Shruti Bhosale},
  {et~al\mbox{.}}} \bibinfo{year}{2023}\natexlab{}.
\newblock \bibinfo{title}{Llama 2: Open foundation and fine-tuned chat models}.
\newblock
\newblock
\showeprint{2307.09288}


\bibitem[Tufano et~al\mbox{.}(2022)]%
        {tufano2022using}
\bibfield{author}{\bibinfo{person}{Rosalia Tufano}, \bibinfo{person}{Simone
  Masiero}, \bibinfo{person}{Antonio Mastropaolo}, \bibinfo{person}{Luca
  Pascarella}, \bibinfo{person}{Denys Poshyvanyk}, {and}
  \bibinfo{person}{Gabriele Bavota}.} \bibinfo{year}{2022}\natexlab{}.
\newblock \bibinfo{title}{Using pre-trained models to boost code review
  automation}.
\newblock
\newblock
\showeprint{2201.06850}


\bibitem[Wang et~al\mbox{.}(2022)]%
        {wang2022recode}
\bibfield{author}{\bibinfo{person}{Shiqi Wang}, \bibinfo{person}{Zheng Li},
  \bibinfo{person}{Haifeng Qian}, \bibinfo{person}{Chenghao Yang},
  \bibinfo{person}{Zijian Wang}, \bibinfo{person}{Mingyue Shang},
  \bibinfo{person}{Varun Kumar}, \bibinfo{person}{Samson Tan},
  \bibinfo{person}{Baishakhi Ray}, \bibinfo{person}{Parminder Bhatia},
  {et~al\mbox{.}}} \bibinfo{year}{2022}\natexlab{}.
\newblock \bibinfo{title}{ReCode: Robustness Evaluation of Code Generation
  Models}.
\newblock
\newblock
\showeprint{2212.10264}


\end{thebibliography}

\end{document}